\documentstyle[amsfonts,aps,epsfig,twocolumn]{revtex}
\newcommand {\be}{\begin{equation}}
\newcommand {\ee}{\end{equation}}
\newcommand{\bea}{\begin{eqnarray}}
\newcommand{\eea}{\end{eqnarray}}
\newcommand{\ba}{\begin{array}}
\newcommand{\ea}{\end{array}}

\newcommand{\sech}{{\rm sech}}
\newcommand{\F}{{\cal F}}

\begin{document}

\title{Strain-induced kinetics of intergrain defects 
as the mechanism of slow dynamics in the 
nonlinear resonant response of humid sandstone bars}

\author{Oleksiy~O.~Vakhnenko~$^1$, Vyacheslav~O.~Vakhnenko~$^2$,
Thomas~J.~Shankland~$^3$, and  James~A.~Ten Cate~$^3$}

\address{$^1$
Bogolyubov Institute for Theoretical Physics, 14-B Metrologichna 
Str., Ky\"\i v 03143, Ukraine \\ 
$^2$ Institute of Geophysics,  63-B Bohdan 
Khmel'nyts'kyy Str., Ky\"\i v 01054, Ukraine \\ 
$^3$ Los Alamos 
National Laboratory, Earth and Environment Sciences Division, Los 
Alamos, New Mexico 87545}

\wideabs{
\maketitle

\begin{abstract}
A closed-form  description is proposed to explain nonlinear and slow 
dynamics effects exhibited by sandstone bars in longitudinal
resonance experiments. Along with the fast subsystem of 
longitudinal nonlinear displacements we examine the strain-dependent 
slow subsystem of broken intergrain and interlamina 
cohesive bonds. We show that even the simplest but 
phenomenologically correct modelling of their mutual feedback 
elucidates the main experimental findings typical for forced 
longitudinal oscillations of sandstone bars, namely, (i) 
hysteretic behavior of a resonance curve on both its up- and 
down-slopes, (ii) linear softening of resonant frequency with 
increase of driving level, and (iii) gradual recovery (increase) 
of resonant frequency at low dynamical strains after the sample 
was conditioned by  high strains. In order to reproduce the 
highly nonlinear elastic features of sandstone grained structure a 
realistic non-perturbative form of strain potential energy was 
adopted. In our theory slow dynamics associated with the 
experimentally observed memory of peak strain history is 
attributed to strain-induced kinetic changes in 
concentration of ruptured inter-grain and inter-lamina cohesive 
bonds causing a net hysteretic effect on the elastic Young's 
modulus. Finally, we explain how  enhancement  of hysteretic 
phenomena originates from an increase in 
equilibrium concentration of ruptured cohesive bonds that are due to water 
saturation.   
\end{abstract}
}

\vspace{0.5cm}

Apart from their excellent static characteristics as building
materials, sandstones have been shown to demonstrate a number of
unexpected and even surprising dynamical properties [1--5]. Here 
we consider the numerous  experimental results on nonlinear 
resonant response exhibited by sandstone rods in forced 
longitudinal oscillations that appear even at extremely small 
forcing levels and consequently at small dynamic strains [1--5]. 
The most intriguing nonlinear feature is slow dynamics, which is 
defined here as long-term (minutes to hours) change of elastic 
properties in response to a disturbance such as dynamic and 
static strain or temperature. 

Specifically, we have to underline that in the vicinity of bar 
resonant frequency the longitudinal alternating drive  
produces strong essentially non-trivial nonlinear  responses: 1) 
At high drive levels the effective width of resonance curves 
depends on the direction of frequency sweep; it is narrower  for 
upward sweeps (i.e., from lower to higher frequencies) than at 
downward sweeps (i.e., from higher to lower frequencies) [1--5].  
This effect proves to be a typical manifestation of slow dynamics 
and can be treated as hysteresis both on low- and high-frequency 
slopes of a resonance curve. 2) The resonance peak is shifted 
toward  lower frequency almost linearly with increase of driving 
amplitude [1, 4].  3) Other evidence of slow dynamics comprises 
gradual recovery (increase) of resonant frequency to its original 
value as defined at extremely low drive level after the sample 
has been conditioned at high drive level [3,5].

These facts cannot be understood in the framework of standard
theories of resonant nonlinear response [6] and imply memory of 
peak strain history [2].  Some aspects of the problem have been 
explained  by the interpretation of Guyer, McCall and Van Den 
Abeele [7] in the framework of a McCall-Guyer quasistatic 
model [8]. This approach uses the concept of 
auxiliary hysteretic elements that allow the introduction of an 
additional nontrivial nonlinear term into the dynamical equation 
for the field of longitudinal displacements.  However, this 
theoretical treatment lacks completeness 
in that it initially neglects the dynamics of hysteretic 
elements and postulates temporal 
evolution of amplitude-frequency characteristic (the key point of 
claimed results) to be developed afterwards.  
Although Capogrosso-Sansone and 
Guyer recently suggested a dynamical realization of the 
McCall-Guyer quasistatic model [9], evaluating its adequacy to 
explain experimental data turns out to be difficult.

In this communication we omit the idea of auxiliary 
hysteretic elements as the sole approach for treating all peculiar 
hysteretic phenomena and call attention to an alternative 
notion used by Davydov and Ermakov for the description of 
bistability in nonlinear resonant tunneling of electrons through 
a set of potential barriers [10]. Their approach consists of 
explicit but physically motivated separation of given 
physical system into two nonlinear subsystems, namely fast and 
slow subsystems with mutual coupling taken into account. 

For sandstone bars we identify the fast subsystem with the 
field of rapid longitudinal displacements while the slow 
subsystem represents  the concentration of defects in intergrain 
contact bonds. In doing this we bear in mind that, because of 
preferable vapor condensation onto surfaces with greater concave 
curvature [11], the sandstone pore structure [4, 11] retains some 
residual pore water [11], and its impact on the resonant 
properties of rock is crucial [12, 13]. Thus, thermodynamical 
estimations applied to porous rocks show that intergrain cohesive 
forces become weaker in the  presence of water [14], that agrees 
with an alternative conception of swelling pressure [13, 15]. 
This treatment is supported by recent experiments [13] 
establishing an abrupt decrease in Young's modulus within the 
first twenty percent interval of water saturation (i.e., until 
the degree that void surfaces become completely wet). We could 
additionally invoke ordinary capillary forces [13] or hydrolysis 
of silicon-oxygen-silicon bond-chains [16] in our consideration.  
However, either of these mechanisms also leads to softening 
of Young's modulus with saturation increase.  Here the significant 
issue is apparently not in excessive (presumably unclaimed) 
detailing all plausible mechanisms that 
might modify the Young's modulus in a qualitatively similar way,  
but in their reasonable concise formalization by means of a 
minimal number of slow fields.    

According to Kosevich [17] the equilibrium concentration of 
defects associated with a stress $\sigma$ is given by the formula 
\begin{equation} 
c_\sigma=c_0\exp\left(v\sigma/kT\right), 
\end{equation} 
where $k$ and $T$ are  the Boltzmann constant and the absolute 
temperature, respectively, and the parameter $v>0$ stands for 
a typical volume accounting for a single defect and 
characterizes the intensity of dilatation. The equilibrium 
concentration of defects in an unstrained bar $c_0$ has to be 
some function of both temperature $T$ and water 
saturation $s$. In order to describe strain-induced changes 
in nonequilibrium concentrations of defects $c$, we assume that at 
any instant of time $t$ the concentration $c$ must evolve to its 
would-be equilibrium value $c_\sigma$, where the stress 
$\sigma$ in (1) is applied at the same instant. Supposing the 
distributions of activation barriers for defect annihilation 
$U$ and activation barriers for defect creation $W$ to be 
uniform respectively  over the ranges $U_0\leq U\leq U_0+U_+$ 
and $W_0\leq W\leq W_0+W_+$ with $U_0$, $U_+$ and $W_0$, $W_+$ 
being insensitive to the choice of bar cross-section, we will 
deal with the density of defect concentration $g$ governed by the 
following kinetic equation 
\begin{equation} 
\partial g/\partial 
t=-\left[\mu\theta(g-g_\sigma)+ 
\nu\theta(g_\sigma-g)\right](g-g_\sigma).  
\end{equation}
Here $\mu=\mu_0\exp(-U/kT)$ and $\nu=\nu_0\exp(-W/kT)$ are the 
rates of defect annihilation and defect creation respectively, 
$g_\sigma=c_\sigma/U_+W_+$, and $\theta(z)$ designates the 
Heaviside step-function. The quantities $g$ and $c$ are related 
by the simple definition
\begin{equation}
c=\int_{U_0}^{U_0+U_+} dU\, \int_{W_0}^{W_0+W_+}\, dW\cdot g.
\end{equation}
Under the tensile load there is an immense number of spatial ways 
for the intergrain cementation contact to be cleaved with the same 
basic result: the creation of crack. The similar scenario is true 
also for the already existed balanced crack to be further 
expanded. On the contrary under the compressive load the crack 
ones emerged has only one spatial way to be annihilated or 
contracted. These observations are the principal ones and imply 
the huge disparity $\nu_0\gg \mu_0$ between the priming rates 
$\nu_0$ and $\mu_0$ notwithstanding the generic cohesive 
properties of cementation material. Moreover, because of possible 
fragmentation of cementation material and/or water intercalation 
between the opposite faces of crack we can expect the typical 
value of barrier $U$ to exceed that of barrier $W$. In 
combination all these factors might sustain predominantly even 
the more immense disparity $\nu\gg\mu$ between the actual rates 
$\nu$ and $\mu$ of defect creation  and defect annihilation, 
apparently comprising many orders, and as a result provide the 
physical mechanism that breaks the symmetry of system response to 
an alternating external drive and acts as a sort of soft ratchet 
or leaky diode.

To express the evolution equation
\begin{equation}
\rho\frac{\partial^2 u}{\partial t^2}=\frac{\partial 
\sigma}{\partial x}  +\frac{\partial }{\partial x}\left[ \frac{\partial 
\F}{\partial (\partial^2 u/\partial x \partial t)}\right]
\end{equation} 
for the field of longitudinal displacements $u$ we 
choose the stress-strain relation in the form
\begin{eqnarray}
\sigma &=& 
{E\sech\eta\over (r-a)[\cosh \eta\, \partial u/\partial 
x+1]^{a+1}}-\nonumber \\
 &-& {E\sech \eta\over (r-a)[\cosh\eta\, \partial  
u/\partial x+1]^{r+1}}
\end{eqnarray} 
which at $r>a>0$ allows one to block the bar compressibility at strains 
$\partial u/\partial x$ tending to $+0-\sech\eta$. To put it 
differently the parameter $\sech\eta$ is reserved for the typical 
thickness of intergrain cementation contact divided by the 
typical distance between the centers of neighbouring grains.
The dissipative function $\F$ we take in the form
\begin{equation} 
\F=(\gamma/2) \left[\partial^2 u/\partial x\, 
\partial t\right]^2
\end{equation} 
giving rise to Stokes internal friction [18].
Here $x$ denotes the 
longitudinal Lagrange coordinate of the bar sample.  The 
quantities $\rho$ and $\gamma$ are respectively the mean density 
of sandstone and the coefficient of internal friction in an 
elastic subsystem. We ignore their dependences on temperature and 
water saturation assuming that the main effect is manifested 
through the linear decrease of Young's modulus $E$ with the 
concentration of defects 
\begin{equation} 
E=\left(1-c/c_{\rm  cr}\right) E_+.  
\end{equation} 
Here $c_{\rm cr}$ and $E_+$ are 
the critical concentration of defects and the maximal possible 
value of Young's modulus, respectively.  Both of these parameters 
we also take to be independent of temperature and water 
saturation.

Typical resonant response experiments [1--5] correspond to 
the kinematic excitation [19] of a bar sample, which we associate 
with the following boundary conditions 
\begin{eqnarray} 
u(x=0|t)&=&  D(t)\cos
\left(\varphi+\int_0^t \, d\tau\omega(\tau) \right)\\ 
\frac{\partial u}{\partial x}(x=L|t) &=& 0
\end{eqnarray}
where $L$ is the sample length and $t>0$. The driving amplitude $D(t)$
is assumed to be basically constant except for the moments when the
driving system is switched on, is switched into another constant
driving level, or is switched off. The time dependence of cyclic driving
frequency $\omega(t)$ in turn is prescribed by the chosen regime of
frequency sweep. Initial conditions are given in the form 
\begin{eqnarray} 
u(x|t=0) &=& 0, \quad \frac{\partial 
u}{\partial t}(x|t=0)=0 \qquad\, (0<x<L) \\ g(x|t=0) &=& c_0/U_+ 
W_+  \qquad  \qquad \qquad\quad  (0<x<L).  \end{eqnarray}

When experimental data for Young's modulus in 
unstrained samples are obtainable from the resonant response 
experiments by the use of a low amplitude protocol (driving amplitude $D$ 
very small and negligible strain-induced 
defects), we can compare them with 
values taken from equation (7) at $c=c_0$ in order to fit 
the equilibrium concentration of defects $c_0$ as a function of 
$T$ and $s$ by some extrapolation  formula. In particular, 
relying upon the Sutherland temperature extrapolation [20] and 
analyzing temperature dependent data at zero saturation [21] 
and saturation dependent data at room temperature [13] 
for Berea sandstone, we suggest the formula 
\begin{equation} 
c_0=c_{\rm cr} \left({T\over T_{\rm 
cr}}\right)^2 \left[\cosh^2\alpha-\exp\left(-{\beta s\over 
1-s}\right)\sinh^2\alpha\right]
\end{equation}
with the following fitting parameters $T_{\rm cr}$=1475 K,  
$\cosh^2\alpha=16$, $\beta=10$. Here the saturation can vary 
within the interval $0\leq s\leq1$. At $s\neq0$ this 
approximation is expected to work at least for temperatures 
exceeding the freezing-point of pore water.

Computer modelling of nonlinear and slow dynamics effects 
was performed in the vicinity of the resonant
frequency $f_0(2)$, which we understand as the second frequency
$(l=2)$ in the fundamental set
\begin{equation}
f_0(l)=\frac{2l-1}{4L} \sqrt{\left(1-{c_0\over c_{\rm 
cr}}\right){E_+\over \rho}} \qquad (l=1, 2, 3, \ldots )
\end{equation}
given by the linear theory of kinematic excitation 
for zero dissipation $\gamma=0$.

Figure 1 shows typical resonance curves, i.e. dependences of 
response amplitude $R$ (calculated at $x=L$) on drive frequency 
$f=\omega/2\pi$, at successively higher drive amplitudes $D$. The 
continuous lines correspond to the conditioned resonance curves 
calculated after two frequency sweeps were performed at each 
driving level in order to achieve repeatable hysteretic curves. 
The dotted line illustrates an unconditioned curve obtained 
without any preliminary conditioning.  Arrows on the three 
highest curves indicate sweep directions.  For the sake of 
definiteness the results of the computer simulation were adapted
to the experimental conditions for the data 
obtained by Ten Cate and Shankland for Berea sandstone [2].  In 
particular, the ratio $E_+/\rho$ was estimated by means of 
relationships (13) and (12) with the second order frequency, bar 
length, temperature and saturation as follows:  $f_0(2)$=3920Hz, 
$L=0.3$m, $T=297$ K and $s=0.25$.  The ratio  $\gamma/\rho$ 
characterizing internal friction was chosen from the best fit to 
low amplitude theoretical (Figure 1) and experimental [2] 
resonance curves using the quality factor $Q$ from resonance 
lineshape. The parameters  $\mu_0\exp(-U_0/kT)=1{\rm s}^{-1}$ 
and $U_+/k=2525$ K determining the character  of slow 
relaxation were estimated according to the experimental 
measurements of decay of acceleration at fixed frequency [2] and 
the observations of recovering resonant frequency 
as a function of time [5]. The combination of 
parameters $vE_+/k\cosh\eta=275$~K was chosen to 
quantitatively reproduce the hysteretic phenomena in the sweep regimes 
typical for the actual experiments [2].  The nonlinearity parameter 
$\cosh\eta=2300$ was estimated to map 
the true asymmetry of experimental resonance curves [2]. 
Other parameters appearing  in the stress-strain relation (5) 
have been adopted as follows: $r=4$, $a=2$.
\begin{figure}[ph]
\centering \includegraphics[height=5.3cm]{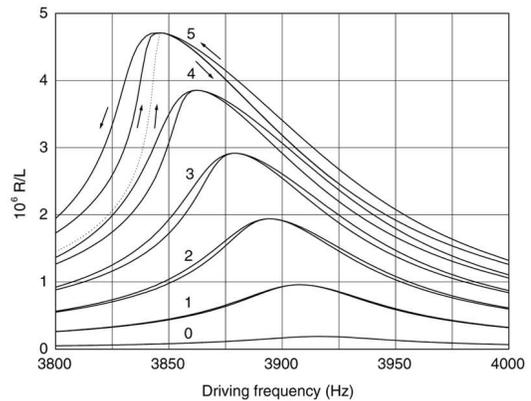}
\caption
{Resonance curves $j=0, 1,  2,  3,  4,  5$ at
 successively higher driving amplitudes $D_j:D_j/L= $ $=
3.8[j(1-\delta_{j0})+0.5\delta_{j0}]\cdot10^{-8}$. Continuous 
lines denote conditioned curves; the dotted line represents 
the unconditioned curve. Arrows on the highest curves indicate 
sweep direction. The time to sweep back and forth within 
the frequency interval  3700Hz$\div$4100Hz is chosen to be 120 s.}

\end{figure}


We would like to stress that through the drop of equilibrium 
concentration $c_0$  our theory is able to catch the 
drastic suppression of hysteresis  with decrease of 
water saturation (see Eq.~(12)).  This conclusion has been 
confirmed by direct computation (not shown). Simultaneously we 
have observed a monotonic increase in quality factor $Q$ with 
saturation decrease, i.e.,  precisely the well documented 
tendency in experiments [12].

Figure 2 compares the shifts of resonant frequency as functions of
driving amplitude at two essentially different  values of dilatation 
parameter $v$ while other parameters were kept the same as 
for Figure 1. Thus curve 1  
calculated at $vE_+/k\cosh\eta=275$~K, when the 
strain-induced feedback between the slow 
and the fast subsystems is substantial, demonstrates the almost 
linear dependence typical for materials with nonclassical 
nonlinear response, i.e.,  materials possessing all the basic 
features of slow dynamics. On the other hand, curve 2 calculated 
at $v=0$, when the strain-induced excitation of slow subsystem is 
absent and hence the mutual feedback between the 
slow and the fast subsystems is totally broken, demonstrates the 
almost quadratic dependence typical for the materials with 
classical nonlinear response. 
\begin{figure}[ph]
\centering \includegraphics[height=5.3cm]{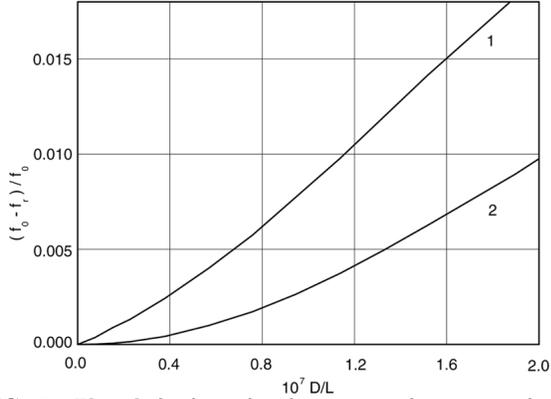}
\caption
{The shift $f_r-f_0$ of resonant frequency $f_r$ from its
 asymptotic value $f_0$ as the function of driving amplitude $D$
 for the hysteretic nonlinear material (curve~1)  and for the classical 
nonlinear material with $v=0$ (curve~2). }
\end{figure}

\begin{figure}[ph]
\centering \includegraphics[height=5.3cm]{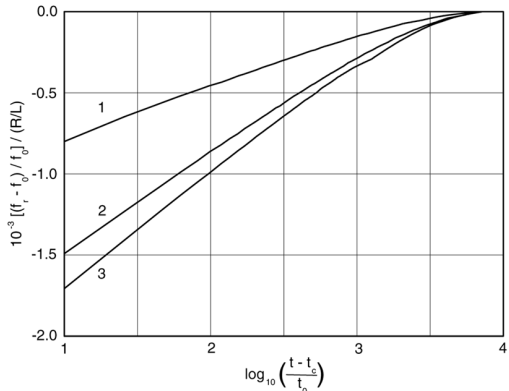}
\caption
{
Time-dependent recovery of resonant frequency $f_r$
 to its asymptotic value $f_0$ after the large conditioning drive
 has been removed. Curves 1, 2, 3  correspond to
 successively higher saturations $s=0.05$, $s=0.15$, $s=0.25$.  
The frequency shift $f_r-f_0$ is normalized by both the 
asymptotic frequency $f_0$ and the unitless response  
amplitude $R/L$ attained at conditioning resonance.}

\end{figure}

Finally, Figure 3 shows the gradual recovery of resonant frequency
$f_r$ to its maximum limiting value $f_0$ after the bar has been
subjected to high amplitude conditioning and conditioning is stopped. 
In the computer simulation we have plotted three 
different curves corresponding to three different saturations 
with all other  model parameters adopted earlier for Figure 
1  being preserved.  The total shift of resonant 
frequency $f_r-f_0$ consists of two physically different parts, 
namely (i) the traditional dynamic shift caused by strain 
nonlinearity at high levels of excitation and (ii) the shift 
caused by the effect of the slow subsystem. However, only the second 
part might actually be registered during the recovery process, 
because the first one vanishes almost instantaneously when the 
conditioning was switched off. Hence, the whole character of 
recovery should inevitably be governed by the slow kinetics 
responsible for restoration of intergrain bonds [5]. 
From Figure 3 we clearly see that the suggested kinetic equation 
for the density of defect  concentration  (2)  supplemented by 
the simple definition of total concentration (3) and the 
reasonable relationship between Young's modulus and a 
concentration of defects (7) yields the very wide time interval 
$10\leq(t-t_{\rm c})/t_0\leq1000$ of logarithmic recovery of the 
resonant frequency in complete agreement with experimental 
results [5]. Here $t_{\rm c}$ is the moment when conditioning 
switches off and $t_0$=1 s. 

This work was carried out within the framework of project No~1747
supported by the STCU. O.O.V.
acknowledges support from the National Academy of Sciences of
Ukraine (Grant 0102U002332). We would like to thank Paul A. 
Johnson for his unflagging interest to the process of multi-stage 
amendments of the model.

\end{document}